\documentclass{article}
\usepackage{spconf,amsmath,graphicx}
\usepackage{mathtools,array,booktabs}  

\title{COVID-19 PATIENT DETECTION FROM  TELEPHONE QUALITY SPEECH DATA}
\name{Kotra Venkata Sai Ritwik, Shareef Babu Kalluri,  Deepu Vijayasenan }
\address{Department of Electronics and Communication Engineering,\\ National Institute of Technology Karnataka - Surathkal, Mangalore, India}
\usepackage{color}

\begin{document}

%
\maketitle
\begin{abstract}
In this paper, we try to investigate the presence of cues about the COVID-19 disease in the speech data.   We use an approach that is similar to speaker recognition. Each sentence is represented as super vectors of short term Mel filter bank features for each phoneme.  These features are used to learn a two-class classifier to separate the COVID-19 speech from normal. Experiments on a small dataset collected from YouTube videos show that an SVM classifier on this dataset is able to achieve an accuracy of 88.6\% and an F1-Score of 92.7\%.  Further investigation reveals that some phone classes, such as nasals, stops, and mid vowels can distinguish the two classes better than the others. 
\end{abstract}
\section{Introduction}
\label{sec:intro}
The pandemic, due to the novel coronavirus (COVID-19), has spread around the world.  In such a situation, the identification of people infected with COVID-19 is challenging for health organizations as well 
as individuals. Major symptoms such as a rise in body temperature, cough, difficulty in breathing are observed; however, totally asymptomatic cases are also possible\cite{WHO}. The clinical protocols in identifying whether the individual is infected with coronavirus include swab test \cite{CDC}, CT scans \cite{gozes2020rapid}, chest X-Ray Images \cite{wang2020covid} etc. In this context, leveraging bio-markers like speech and audio signals in screening the COVID-19  will help in the assessment of the viral infection. Different studies and pathological investigations have proved that COVID-19 infected individuals exhibit difficulty while breathing and speaking. Furthermore, the changes in speech might not be identifiable by human perception, although the person is infected, but the Computer auditions (CA) \cite{quatieri2020framework,schuller2020covid} can. As a token of contribution to the society in this pandemic situation, we tried to investigate on detection of the COVID-19 infection from speech sounds alone.   In this process, we also focus on telephone speech so that such a system could be integrated to assess the risk of the pandemic at a region by telephone operators. 

There are a few attempts in the literature in identifying pathological conditions such as bronchitis and pertussis using mainly cough patterns  \cite{windmon2018detecting}.   These research efforts perform machine learning techniques on speech samples collected from smartphone recordings. From the studies on the COVID-19 patients, it was noticed that the respiration activity would be more rapid for the infected people than other patients with flu and common cold \cite{wang2020abnormal}. Other studies show that the respiratory parameters are also have impact on the persons mood/emotion\cite{hameed2019human}, stress levels \cite{perciavalle2017role} and physiological state of individuals \cite{griffiths2019guidelines}. In the context of the Interspeech Computational Paralinguistics Challenge (ComParE), different research efforts have attempted to estimate the behavioral state of humans using speech data in conditions like cold, cough, pain, sleepiness, and infant cries. Schuller et al. has suggested a set of computer audition tasks using speech analysis and sound analysis for COVID-19 risk assessment using machine learning techniques \cite{schuller2020covid}.   However, in this work, we look only at the speech signal and not at breathing or cough patterns that may be difficult to acquire through a noisy telephone channel.   Also, in this case, the user of the system does not have to force a cough or create breathing patterns. 
To the best of the authors' knowledge, this is the first work on COVID-19 screening using only telephone quality speech data. 

The organization of the paper is as follows, Section \ref{sec:feat} details the feature extraction and statistical representation of each phoneme as well as sentence. It also details the classification methods. In Section \ref{sec:dataset}, the details of the collected dataset are briefed. Later the experiments and results are detailed in Section~\ref{sec:expts}. Finally, the conclusions of the reported work and future directions of the proposed approach are presented in Section \ref{sec:concl}. 

\section{Approach}
\label{sec:feat}
\begin{figure}
    \centering
    \includegraphics[width = \linewidth]{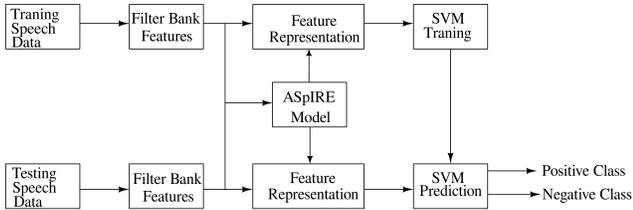}
    \caption{Block diagram for detection of COVID-19 positive or negative class using speech data}
    \label{fig:blockdia}
\end{figure}

Many attempts have been made to automate the disease identification using the respiratory patterns \cite{wang2020abnormal}, cough patterns \cite{windmon2018detecting} and also COVID-19 \cite{quatieri2020framework}. These attempts make use of the bio-markers from speech such as fundamental frequency, short-term cepstrum,  cepstral peak prominence, Harmonics to noise ratio, Glottal open quotient, etc. Higher-level feature derived features from such features were employed to screen the people for COVID-19 \cite{quatieri2020framework}.  In another attempt, a pool of such features across different sounds was used to identify the presence of COVID-19 infection \cite{sharma2020coswara}.  In this work, as well, we use short term mel-spectrum as the low-level features.   We follow an approach similar to super vectors that were originally used in speaker recognition~\cite{kinnunen2010overview} to extract utterance level features.  An SVM  trained with this feature input is used to predict COVID-19 from speech data.   The block diagram of the approach is shown in Figure \ref{fig:blockdia}.  We could not 
use the more advanced version of super-vectors such as the i-vectors~\cite{dehak2010front} or x-vectors~\cite{snyder2018x}  due to limited amount of training data. 

\subsection{Super vector Features}
Traditionally super vectors are extracted using the posterior probabilities of a 
background Gaussian Mixture Model.  However, the dataset we worked on was from YouTube videos and was very noisy.  Therefore we decided to use a deep neural network model as the phoneme posterior probability estimator rather than a GMM.  We used the ASpIRE chain model~\cite{ko2015audio}.  

The model is a Time-Delayed Neural Network(TDNN) 
that is trained on the Fisher English dataset.   Data augmentation by means of different room impulse responses and noises are incorporated to increase the robustness~\cite{ko2017study}.  The model outputs the frame-level posteriors of different context-dependent phonemes. 

Consider an input sequence of short term mel-spectral features from a speech utterance $\{\bf x_, x_2, \ldots, x_T\}$. The Aspire model takes the short term mel-cepstrum as the input and predicts the frame level posteriors $\{ \bf p_1,\ldots,p_T\}$  where:
\begin{equation}
{\bf p_j} = \begin{bmatrix}{}
     p^1_j  \\
     p^2_j \\
     \vdots \\
     p^M_j 
\end{bmatrix} 
\end{equation}
with $p^i_j$ denoting the posterior probability of phoneme $i$ at the frame index $j$. 
Once the posterior probabilities are calculated, we compute the normalized first order statistics of each  phoneme as :
\begin{equation}
\label{eqn:fstat}
    {\bf f^i} = \frac{1}{\sum_j p^i_j}\sum_j p^i_j \bf x_j
\end{equation}
 We then concatenate all frames of ${\bf f^i}$ of  each phoneme to obtain a super vector ${\bf F^{i} }= [f_1,f_2, \ldots ,f_j ]$ which represents the utterance level feature.
 We had to resort 
to this primitive approach as there was no large scale development data available in this domain for learning more advanced feature representations such as i-vectors or x-vectors. 

\subsection{Classification}
 The computed statistics ($\bf F^{i}$) are used as features for the learning of a support vector machine(SVM) model for the task.   We have generated multiple training examples from the same sentence by splitting it into many overlapping sentences. A radial basis kernel is used to train the SVM model.  Cross-validation experiments in the training data are performed to determine the hyper-parameters, and the results are reported in an independent test set. 
 
\section{Dataset}
\label{sec:dataset}
The dataset consists of audio clips extracted from YouTube videos of TV interviews (often a video call) of COVID-19 positive patients \footnote{ A sample YouTube link used: \textit{https://www.youtube.com/watch?v=yr2GnZo2KiA}}.
The dataset is made of two distinct classes: The COVID-19 Positive speakers class and the COVID-19 Negative speakers class. The video recordings of the COVID-19 Positive speakers are either from the hospital environment or home isolated/quarantine environment.  Most of the audio, therefore, has a lot of background noise.  The COVID-19 Negative speech recordings are also taken from a controlled studio environment while broadcasting over TV channels/YouTube.   

The speech part of the audio was extracted at the 44.1kHz sampling rate. The audio samples were converted into a single channel (mono) wav format and passed through a 300Hz -- 3.4 kHz band-pass filter. This is performed to simulate telephone quality speech.  The audio is finally down-sampled to 8 kHz range.  Audio corresponding to different speakers are segmented manually, and each utterance was annotated as positive and negative class. COVID-19 positive speakers are annotated as the positive class, and host/other audio speakers were annotated as the negative class. 
Speech non-speech separation was performed on each utterance.

We have collected the speech data from nineteen speakers, of which ten speakers are COVID-19 Positive, and nine speakers are COVID-19 Negative\footnote{ The dataset used for this work is available in the following link \textit{https://github.com/shareefbabu/covid$\_$data$\_$telephone$\_$band}}. The composition of male and female speakers in each class are tabulated in Table \ref{tab:dataset_stats} along with the number of utterances per each class and gender. 

\begin{table}
\centering
\caption{Details of collected dataset of Male (M) and Female (F) Speakers}
\begin{tabular}{|c|c|c|c|c|}
\hline
Class & \multicolumn{2}{c|}{Speakers}                           & \multicolumn{2}{c|}{$\#$ Sentences} \\ \cline{2-5} 
& F & M & F & M \\ \hline
 
Positive    & 4 & 6  & 168  & 296\\ \hline
Negative   & 3 & 6 & 44  & 194\\ \hline
\end{tabular}
\label{tab:dataset_stats}
\end{table}

\section{Experiments and Results}
\label{sec:expts}
We divided the collected data into training (12 speakers) and testing (7 speakers) parts. There are 501 utterances for training and 201 utterances for testing.  There is no overlap  speakers in train and test splits.

The performance of the system is assessed with, accuracy and F1-score metrics. The F1-score metric computation is given by the following equation 
\begin{equation}
F1 score = \frac{2PR} {(P + R)} 
\end{equation}
where the Precision (P) and Recall (R) are given by:
\begin{align*}
P &= \frac{TP} {(TP+FP)}     \\
R &= \frac{TP} {(TP+FN)}
\end{align*}
Where TP denotes the true positives, FP the false positives and FN the false negatives. 


We windowed the speech into the usual 25ms long, 10 ms shifted frames, and extracted 40-dimensional mel filter bank coefficients.   In addition,  the frame-level phone posteriors are computed from the ASpIRE chain model.  Even though the model provides context-dependent phoneme probabilities,  we sum over the context-dependent phonemes to get context-independent phoneme posteriors. Thus we get a 39-dimensional posterior vector corresponding to 39 TIMIT phonemes. The silence phoneme is discarded.   The mel-filter bank features and phoneme posteriors are used in the computation of first-order statistics of every phoneme ( Eqtn.~\ref{eqn:fstat} ). Each phoneme first-order statistics is a $40$ dimensional vector.   These are 
concatenated to form a $40\times 39 = 1560$ dimensional super vector.  We first performed a set of cross-validation experiments in the training data
to optimize certain parameters and then performed an actual evaluation on the test data. 

\subsection{Cross Validation Experiments}
Initially, we performed cross-validation experiments to assess the 
performance, as well as tune some hyper-parameters of the SVM classifier. The training dataset of 12 speakers is divided into six subsets(folds) of 2 speakers each for cross-validation. The training split has six speakers, each in positive (307 utterances) and negative (194 utterances) classes. There are no overlapping speakers across different subsets. Each time, one subset was kept for testing, and the remaining five subsets were used for training the SVM model.

However, the training data was not sufficient in this setting, and we have observed an over-fit problem with the SVM.  To compensate for the same, 
we computed utterance level features using 3s segments from the training data.   The 3s segments were chosen with a shift of 0.1 seconds. This allowed us to increase the number of training examples.   However, testing examples were kept as such. 

The SVM kernel and the regularization parameter were selected based on the maximum accuracy of the cross-validation experiment. 
The Radial Basis Function kernel was found to be the optimal kernel for the SVM.   The best weighted average accuracy and F1 score over all the folds in cross-validation are  70.5\% and 77.0\%, respectively.

To get an understanding of the results, we also conducted an experiment
to assess the contribution of different phoneme classes in COVID-19 detection. This experiment is conducted by using different feature subsets that correspond to different speech classes. 

We performed the same set of cross-validation experiments by considering only a subset of phoneme posteriors. We considered eight different classes of phonemes: nasals, back-vowels, front-vowels, mid-vowels, semi-vowels, stops, fricatives, and diphthongs.   Each time only posteriors that fall in the sound class were used to compute the first-order statistics(refer Eqtn:~\ref{eqn:fstat}). For example, to run the experiments only for nasal phoneme class we will 
only use three posteriors (corresponding to /m/, /n/ and /ng/ ). Therefore the utterance lever feature will be only $3\times40 = 120$ dimensional. In order to avoid sentences that do not have these phonemes, we put a threshold value on the sum of the posteriors across all frames and within the phone class.  This threshold value was chosen as $30$ based on cross-validation experiments. Only if the value is greater than the threshold, the sentence was considered for testing.   If we assume that the DNN was predicting all labels correctly, this is equivalent to saying that we expect at least 
30 frames within the phoneme class (say nasals) in the sentence.

\begin{figure}[!h]
    \centering
\includegraphics[width=\linewidth]{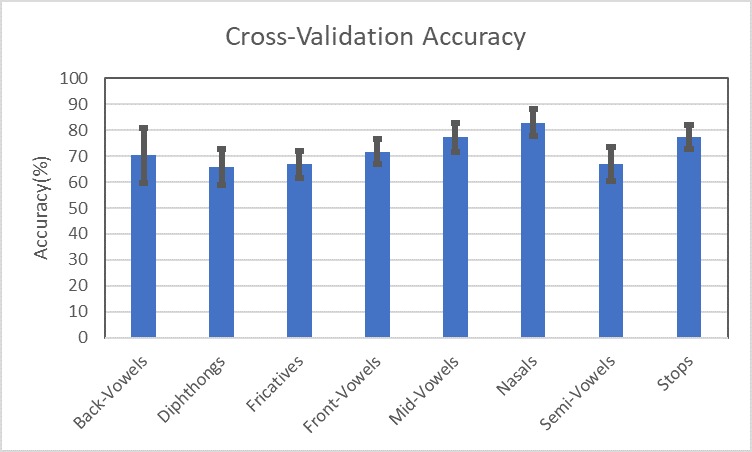}
    \caption{Accuracy with confidence interval of 95\% marked for specific phonemes on cross-validation data}
    \label{fig:phn_cross}
\end{figure}
\vspace{-10pt}
\begin{figure}[!h]
    \centering
\includegraphics[width=\linewidth]{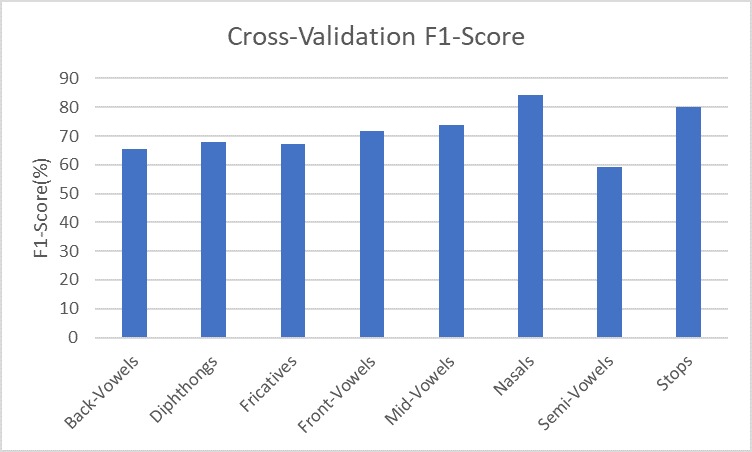}
    \caption{F1 scores for specific phonemes on cross-validation data}
    \label{fig:phnf1_cross}
\end{figure}
Fig.\ref{fig:phn_cross} shows the accuracy  of each class with confidence interval of 95\% on cross-validation data. F1 scores for each of the phoneme class are shown in Fig.\ref{fig:phnf1_cross}.

Nasals, Stops, and Mid-Vowels are the top-3 performing classes on cross-validation with accuracies of 82.9\%, 77.52\%, and 77.25\%, respectively, over all the folds. The corresponding  F1-scores for Nasals, Stops, and Mid-Vowels are 84.06\%, 80.23\%, and 73.91\%, respectively. The specificity and sensitivity scores on the cross-validation dataset for phonemes as well as full dataset is given in Table.\ref{tab:sensi_speci_cross}.

\begin{table}[]
    \centering
        \caption{Sensitivity and Specificity values on the cross validation data}
    \begin{tabular}{c|c|c} \noalign{\smallskip}\hline
& Specificity & Sensitivity \\ \hline
Full Dataset & 0.55 & 0.81  \\
Nasals       & 0.80 & 0.85 \\
Front-Vowels & 0.73 & 0.70 \\
Mid-Vowels   & 0.81 & 0.73 \\
Back-Vowels  & 0.70 & 0.71 \\
Semi-Vowels  & 0.74 & 0.58 \\
Stops        & 0.67 & 0.87 \\
Fricatives   & 0.67 & 0.67 \\
Diphthongs   & 0.62 & 0.71 \\ \hline
    \end{tabular}

    \label{tab:sensi_speci_cross}
\end{table}


\subsection{ Evaluation data }
After fixing all hyper-parameters of the system, we evaluate the system
performance on independent test data.  The SVM is trained using the entire training data using optimized parameters from the cross-validation experiment. The system is evaluated on the 7 speakers test dataset (201 utterances) out of which 4 speakers are in positive class (157 utterances) and 3 speakers are in negative class (44 utterances). 

\begin{figure}[!h]
    \centering
\includegraphics[width=\linewidth]{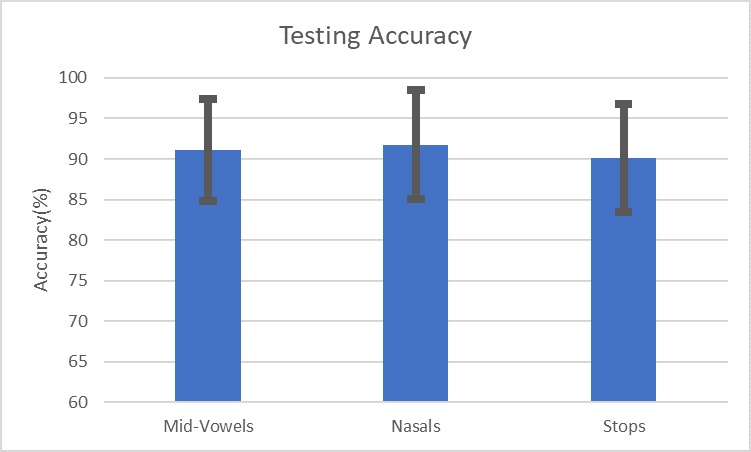}
    \caption{Accuracy with confidence interval of 95\% marked for specific phonemes on test data}
    \label{fig:phn_test}
\end{figure}

\begin{figure}[!h]
    \centering
\includegraphics[width=\linewidth]{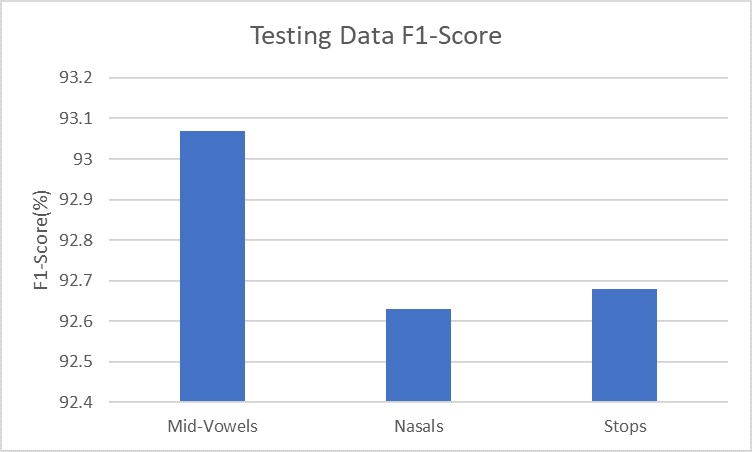}
    \caption{F1 scores for specific phonemes on test data}
    \label{fig:phnf1_test}
\end{figure}
The accuracy and F1 score over the testing data were found to be 88.6\% and 92.7\%, respectively.  It may be noted that the system is able to generalize the performance of the cross-validation experiment to the test set. The ROC 
curve on the evaluation data is shown in Fig~\ref{fig:full_roc}.  A very recent cough-based system~\cite{imran2020ai4covid} has reported accuracy of $88.9\%$ and an F-Score of $90.3\%$ on identifying $4$ different pathological conditions.  However, the system relies on the cough signal, that is not as easily available as the speech signal. Besides, our system has the advantage of telephone quality speech data that can be acquired remotely.  

We also tested the system using only phoneme sub-classes.   The highest performing sound classes are the same -- nasals, stops, and mid-vowels with accuracies { 91.8\%, 90.1\%, and 91.1\%} and the corresponding F1-Scores are {92.6\%, 92.7\% and 93.1\% respectively.   The specificity and sensitivity scores on the evaluation dataset for phonemes  is given in Table.\ref{tab:sensi_speci_eval}.  The corresponding ROC curves are shown in Figure~\ref{fig:roc_phn}.  This setup shows a better performance compared to using all phoneme classes together as feature input in terms
of accuracy, F1 score, and area under the curve (AUC). Thus it could be concluded that these phoneme classes carry some bio-markers about the COVID-19 infection.    However, 
it may be noted that the test dataset in phoneme specific classes is only a subset
of the total subset because all the phoneme sub-classes do not occur in every sentence.    We noticed that only around $40\%$ of the total test set had each of these phoneme classes(nasals, mid-vowels, and stops). 

\begin{table}[]
    \centering    
    \caption{Sensitivity and Specificity values on the Test data}
        \label{tab:sensi_speci_eval}
    \begin{tabular}{c|c|c}  \noalign{\smallskip}\hline
& Specificity &  Sensitivity \\ \hline
Full Dataset  &  0.73 &  0.93 \\
Nasals        &  0.85 &  0.94 \\
Mid-Vowels    &  0.89 &  0.92 \\
Stops         &  0.82 &  0.90 \\ \hline
    \end{tabular}
\end{table}


\begin{figure}[t]
    \centering
\includegraphics[width=\linewidth]{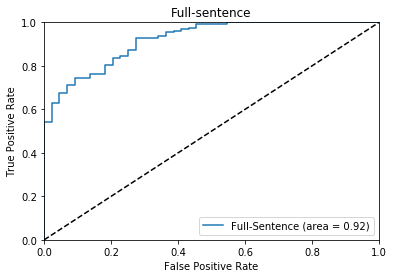}
    \caption{ROC Curve for COVID detection with features from all phonemes (full sentence)}
    \label{fig:full_roc}
\end{figure}

\begin{figure}[t]
    \centering
\includegraphics[width=\linewidth]{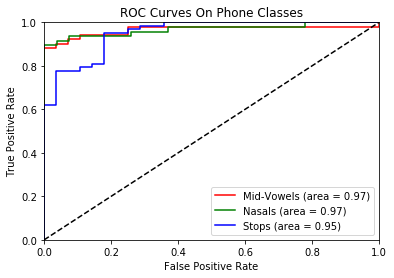}
    \caption{ROC Curve for COVID detection with features for three top performing phoneme classes}
    \label{fig:roc_phn}
\end{figure}

\section{Conclusion}
\label{sec:concl}
We created a dataset of speech from COVID-19 positive speakers.  We used 
primitive super-vector like features as utterance level features due to data size limitations.   SVM based machine learning systems achieved an accuracy of 88.6\% with an F-score of 92.7\%.   Testing on individual phoneme sub classes shows that Nasals, Stops, and Mid-Vowels alone can predict the presence of COVID-19 from speech data.  The entire study depends only on telephonic quality speech.  
In addition to remote screening, such a system could help predict
the pandemic risk in a region by sampling voice calls.

The study is currently limited to $19$ for speakers only.   The authors would like to acquire more data.  This would enable us to validate our results on the phoneme sub-class level quite extensively.  We hope this will enable us to design a proper telephone-based COVID-screening test in the future.



\bibliographystyle{IEEEbib}
\bibliography{strings,refs}

\begin{thebibliography}{10}

\bibitem{WHO}
World~Health Organization,
\newblock {\em Coronavirus disease (COVID-19) outbreak situation}, 2020.

\bibitem{CDC}
Centers for Disease~Control and Prevention,
\newblock ``{Coronavirus disease (COVID-19) Test for Current Infection},''
  2020.

\bibitem{gozes2020rapid}
Ophir Gozes, Maayan Frid-Adar, Hayit Greenspan, Patrick~D Browning, Huangqi
  Zhang, Wenbin Ji, Adam Bernheim, and Eliot Siegel,
\newblock ``Rapid {AI} development cycle for the coronavirus (covid-19)
  pandemic: Initial results for automated detection \& patient monitoring using
  deep learning ct image analysis,''
\newblock {\em arXiv preprint arXiv:2003.05037}, 2020.

\bibitem{wang2020covid}
Linda Wang and Alexander Wong,
\newblock ``Covid-net: A tailored deep convolutional neural network design for
  detection of covid-19 cases from chest x-ray images,''
\newblock {\em arXiv preprint arXiv:2003.09871}, 2020.

\bibitem{quatieri2020framework}
Thomas~F Quatieri, Tanya Talkar, and Jeffrey~S Palmer,
\newblock ``A framework for biomarkers of covid-19 based on coordination of
  speech-production subsystems,''
\newblock {\em IEEE Open Journal of Engineering in Medicine and Biology}, vol.
  1, pp. 203--206, 2020.

\bibitem{schuller2020covid}
Bj{\"o}rn~W Schuller, Dagmar~M Schuller, Kun Qian, Juan Liu, Huaiyuan Zheng,
  and Xiao Li,
\newblock ``Covid-19 and computer audition: An overview on what speech \& sound
  analysis could contribute in the sars-cov-2 corona crisis,''
\newblock {\em arXiv preprint arXiv:2003.11117}, 2020.

\bibitem{windmon2018detecting}
Anthony Windmon, Mona Minakshi, Sriram Chellappan, Ponrathi Athilingam, Marcia
  Johansson, and Bradlee~A Jenkins,
\newblock ``On detecting chronic obstructive pulmonary disease (copd) cough
  using audio signals recorded from smart-phones.,''
\newblock in {\em HEALTHINF}, 2018, pp. 329--338.

\bibitem{wang2020abnormal}
Yunlu Wang, Menghan Hu, Qingli Li, Xiao-Ping Zhang, Guangtao Zhai, and Nan Yao,
\newblock ``Abnormal respiratory patterns classifier may contribute to
  large-scale screening of people infected with covid-19 in an accurate and
  unobtrusive manner,''
\newblock {\em arXiv preprint arXiv:2002.05534}, 2020.

\bibitem{hameed2019human}
Rabab~A Hameed, Mohannad~K Sabir, Mohammed~A Fadhel, Omran Al-Shamma, and Laith
  Alzubaidi,
\newblock ``Human emotion classification based on respiration signal,''
\newblock in {\em Proceedings of the International Conference on Information
  and Communication Technology}, 2019, pp. 239--245.

\bibitem{perciavalle2017role}
Valentina Perciavalle, Marta Blandini, Paola Fecarotta, Andrea Buscemi,
  Donatella Di~Corrado, Luana Bertolo, Fulvia Fichera, and Marinella Coco,
\newblock ``The role of deep breathing on stress,''
\newblock {\em Neurological Sciences}, vol. 38, no. 3, pp. 451--458, 2017.

\bibitem{griffiths2019guidelines}
Mark~JD Griffiths, Danny~Francis McAuley, Gavin~D Perkins, Nicholas Barrett,
  Bronagh Blackwood, Andrew Boyle, Nigel Chee, Bronwen Connolly, Paul Dark,
  Simon Finney, et~al.,
\newblock ``Guidelines on the management of acute respiratory distress
  syndrome,''
\newblock {\em BMJ open respiratory research}, vol. 6, no. 1, pp. e000420,
  2019.

\bibitem{sharma2020coswara}
Neeraj Sharma, Prashant Krishnan, Rohit Kumar, Shreyas Ramoji, Srikanth~Raj
  Chetupalli, Prasanta~Kumar Ghosh, Sriram Ganapathy, et~al.,
\newblock ``Coswara--a database of breathing, cough, and voice sounds for
  covid-19 diagnosis,''
\newblock {\em arXiv preprint arXiv:2005.10548}, 2020.

\bibitem{kinnunen2010overview}
Tomi Kinnunen and Haizhou Li,
\newblock ``An overview of text-independent speaker recognition: From features
  to supervectors,''
\newblock {\em Speech communication}, vol. 52, no. 1, pp. 12--40, 2010.

\bibitem{dehak2010front}
Najim Dehak, Patrick~J Kenny, R{\'e}da Dehak, Pierre Dumouchel, and Pierre
  Ouellet,
\newblock ``Front-end factor analysis for speaker verification,''
\newblock {\em IEEE Transactions on Audio, Speech, and Language Processing},
  vol. 19, no. 4, pp. 788--798, 2010.

\bibitem{snyder2018x}
David Snyder, Daniel Garcia-Romero, Gregory Sell, Daniel Povey, and Sanjeev
  Khudanpur,
\newblock ``X-vectors: Robust dnn embeddings for speaker recognition,''
\newblock in {\em 2018 IEEE International Conference on Acoustics, Speech and
  Signal Processing (ICASSP)}. IEEE, 2018, pp. 5329--5333.

\bibitem{ko2015audio}
Tom Ko, Vijayaditya Peddinti, Daniel Povey, and Sanjeev Khudanpur,
\newblock ``Audio augmentation for speech recognition,''
\newblock in {\em Sixteenth Annual Conference of the International Speech
  Communication Association}, 2015.

\bibitem{ko2017study}
Tom Ko, Vijayaditya Peddinti, Daniel Povey, Michael~L Seltzer, and Sanjeev
  Khudanpur,
\newblock ``A study on data augmentation of reverberant speech for robust
  speech recognition,''
\newblock in {\em 2017 IEEE International Conference on Acoustics, Speech and
  Signal Processing (ICASSP)}. IEEE, 2017, pp. 5220--5224.

\bibitem{imran2020ai4covid}
Ali Imran, Iryna Posokhova, Haneya~N Qureshi, Usama Masood, Sajid Riaz, Kamran
  Ali, Charles~N John, Iftikhar Hussain, and Muhammad Nabeel,
\newblock ``Ai4covid-19: Ai enabled preliminary diagnosis for covid-19 from
  cough samples via an app,''
\newblock {\em Informatics in Medicine Unlocked}, p. 100378, 2020.

\end{thebibliography}

\end{document}